\documentclass[twocolumn,showpacs,preprintnumbers,amsmath,amssymb]{revtex4}

\usepackage{graphicx}
\usepackage{dcolumn}
\usepackage{bm}

\begin{document}

\title{Mass gap for gravity localized on Weyl thick branes}

\author{N. Barbosa--Cendejas$^1$}
\email{nandinii@ifm.umich.mx}
\author{A. Herrera--Aguilar$^2$}%
\email{herrera@ifm.umich.mx}
\author{M.A. Reyes Santos$^1$}%
\email{marco@fisica.ugto.mx}
\author{C. Schubert$^2$}%
\email{schubert@ifm.umich.mx} \affiliation{$^1$Instituto de
F\'{\i}sica, Universidad de Guanajuato, Loma del Bosque 103, Frac.
Lomas del Campestre, C.P. 37150 Le\'{o}n,
Guanajuato, M\'{e}xico.\\
$^2$Instituto de F\'{\i}sica y Matem\'{a}ticas, Universidad
Michoacana de San Nicol\'as de Hidalgo.\\
Edificio C--3, Ciudad Universitaria, C.P. 58040, Morelia,
Michoac\'{a}n, M\'{e}xico. } 

\begin{abstract}
We consider thick brane configurations in a pure geometric Weyl
integrable 5D space time, a non--Riemannian generalization of
Kaluza--Klein (KK) theory involving a geometric scalar field. Thus,
the 5D theory describes gravity coupled to a self--interacting
scalar field which gives rise to the structure of the thick branes.
We continue the study of the properties of a previously found family
of solutions which is smooth at the position of the brane but
involves naked singularities in the fifth dimension. Analyzing their
graviton spectrum, we find that a particularly interesting situation
arises for a special case in which the 4D graviton is separated from
the KK gravitons by a mass gap. The corresponding effective
Schr\"odinger equation has a modified P\"oschl-Teller potential and
can be solved exactly. Apart from the massless 4D graviton, it
contains one massive KK bound state, and the continuum spectrum of
delocalized KK modes. We also discuss the mass hierarchy problem,
and explicitly compute the corrections to Newton's law in the thin
brane limit.

\end{abstract}

\maketitle

\section{
Introduction}

The success of brane world scenarios to address a number of high
energy physics problems \cite{rubakov} and the expectation of
possible experimental evidence for extra dimensions
\cite{ahddd}--\cite{lr} have given rise to various modifications and
generalizations of the initial thin brane models
\cite{gog}--\cite{lr}. Among them we find several types of thick
brane configurations: scalar smooth branes
\cite{dewolfe}--\cite{scalarthickbranes}, tachyonic branes
\cite{branetachyons}, branes on black holes \cite{BlackHoles,wang},
branes with non--minimally coupled scalar fields
\cite{griegos,farakosetal}; within this framework several physical
aspects have been studied like localization of gravity
\cite{csaki}--\cite{bgl} and matter fields \cite{matterloc}, brane
stability \cite{stabilitythickbranes1,stabilitythickbranes2},
critical phenomena \cite{criticalphenomena}, etc.

The model to be considered here is of the class introduced by
\cite{dewolfe}. It describes Weyl gravity coupled to a geometrical
scalar field and constitutes a non--compact generalization of the
Randall--Sundrum (RS) model \cite{rs1,rs2} with the advantage that
the 5D manifold is no longer restricted to be an orbifold and the
branes in the action need not be introduced by hand.

In this paper, we briefly review localization of 4D gravity on
previously constructed thick branes in a Weyl integrable manifold
\cite{bh3}. In Weyl geometry, a scalar field enters in the
definition of the affine connection of the manifold, revealing its
geometrical nature. Weyl integrable manifolds are invariant under
Weyl rescalings (see, e.g., \cite{bh2} for details); when this
invariance is broken, via a self--interacting potential, for
instance, the scalar field transforms into an observable degree of
freedom.

Various classes of thick brane solutions have been discussed in the
literature. In particular, in
\cite{dewolfe,gremm,csaki,Localgravity2} and
\cite{criticalphenomena,branetachyons,farakosetal}, it was proposed
to smooth out the results obtained in the RS model with the aid of a
scalar field. Moreover, the stability of these thick objects was
considered in \cite{stabilitythickbranes1}. All these models have a
continuous KK graviton spectrum starting at zero mass, causing a
need for explaining why such arbitrarily light extra gravitons have
not led to detectable deviations from standard gravity. Although a
number of mechanisms have been proposed how such light extra
gravitons could go unobserved (see, e.g., \cite{lr}), it is
obviously an interesting question whether thick brane solutions
exist which have a mass gap and thereby avoid this problem
altogether. Reconsidering some of the previously found families of
solutions we have found one such case, and the main purpose of this
paper is a detailed study of its properties.

The solution family in question involves a simple trigonometric warp
factor \cite{dewolfe,gremm},\cite{ariasetal}--\cite{matterloc}.
Following \cite{rs1,dewolfe,csaki}, we study linear fluctuations of
the classical metric and cast the wave equation for the transverse
traceless modes in the form of a Schr\"odinger equation. Contrary to
previous studies of this type of solutions, where the potential in
this equation usually went to zero far from the brane, we fix the
free parameter of the model in such a way that it asymptotes to a
positive constant. The corresponding Schr\"odinger equation turns
out to be exactly solvable, being of P\"oschl--Teller type. Apart
from the massless 4D graviton, it has one more bound state,
describing a massive localized graviton, as well as a continuum
tower of delocalized massive modes. Thus it indeed leads to a mass
gap for the graviton spectrum, whose size is determined by the
coupling constant of the scalar field self--interaction. This comes
at a price, however: The curvature for this solution shows a naked
singularity at a finite geodesic distance from the brane.

In this type of model, the massive modes give rise to small
corrections to the Newton's law in 4D flat space time. This opens a
new avenue to obtain theoretical predictions that can, in principle,
be experimentally tested. Since in our special case we have all
graviton wave functions in closed form, we are able to obtain the
exact form of the correction to Newton's constant in the thin brane
limit. Finally, we comment on the possible use of this configuration
for the solution of the mass hierarchy problem along the lines of
\cite{rs1}--\cite{lr}.

\section{
Definition of the model}

Let us start with a pure geometrical Weyl action in five dimensions.
This non--Riemannian generalization of the Kaluza--Klein theory is
given by
\begin{equation}
\label{action} S_5^W =\int_{M_5^W}\frac{d^5x\sqrt{|g|}}{16\pi
G_5}e^{\frac{3}{2}\omega}[R+3\tilde{\xi}(\nabla\omega)^2+6U(\omega)],
\end{equation}
where $M_5^W$ is a Weyl manifold specified by the pair
$(g_{MN},\omega)$ and $\omega$ is a scalar field. The Weylian Ricci
tensor reads
$R_{MN}=\Gamma_{MN,A}^A-\Gamma_{AM,N}^A+\Gamma_{MN}^P\Gamma_{PQ}^Q-\Gamma_{MQ}^P\Gamma_{NP}^Q,$
where
$\Gamma_{MN}^C=\{_{MN}^{\;C}\}-\frac{1}{2}\left(
\omega_{,M}\delta_N^C+\omega_{,N}\delta_M^C-g_{MN}\omega^{,C}\right)$
are the affine connections on $M_5^W$, $\{_{MN}^{\;C}\}$ are the
Christoffel symbols and $M,N=0,1,2,3,5$; $\tilde{\xi}$ is an
arbitrary coupling parameter, and $U(\omega)$ is a self--interacting
potential for the scalar field $\omega$.

By performing a conformal transformation
$\widehat{g}_{MN}=e^{\omega}g_{MN}$, the action (\ref{action}) is
mapped into the Einstein frame
\begin{equation}
\label{Eaction} S_5^R
=\int_{M_5^R}\frac{d^5x\sqrt{|\widehat{g}|}}{16\pi
G_5}[\widehat{R}+3\xi(\widehat{\nabla}\omega)^2+6\widehat{U}(\omega)],
\end{equation}
where the affine connections become Christoffel symbols,
$\xi=\tilde{\xi}-1$, $\widehat{U}=e^{-\omega}U$ and hatted
magnitudes refer to the Einstein frame. The corresponding field
equations read
\begin{eqnarray}
\widehat{R}_{MN}=-\left(3\xi\widehat{\nabla}_M\omega\widehat{\nabla}_N\omega+
2\widehat{U}\widehat{g}_{MN}\right),\nonumber\\
\widehat{\nabla}^2\omega=\frac{1}{\xi}\frac{d\widehat{U}}{d\omega}.
\label{Rfieldeqs}
\end{eqnarray}
In order to get solutions to the theory (\ref{action}) that preserve
4D Poincar\'e invariance we consider the following ansatz
\begin{equation}
\label{line} ds_5^2=e^{2A(y)}\eta_{mn}dx^m dx^n+dy^2,
\end{equation}
where $e^{2A(y)}$ is the warp factor depending on the extra
coordinate $y$, and $m,n=0,1,2,3$. Thus, the field equations
(\ref{Rfieldeqs}) adopt the form \footnote{We have corrected some
sign mistakes in the field equations (\ref{EqsAomega}) that
originated in \cite{ariasetal} and also led to errors in the
preprint version of this paper (arXiv:0709.3552). These errors
affected only the behaviour of the scalar field, however, not the
graviton fluctuation analysis.}
\begin{eqnarray}
\omega''+4A'\omega'+\frac{3}{2}(\omega')^2=\frac{1}{\xi}
\frac{d\widehat{U}}{d\omega}e^{\omega},\nonumber\\
A''+4(A')^2+\frac{3}{2}A'\omega'=\frac{1}{2}\left(4\widehat{U}-\frac{1}{\xi}
\frac{d\widehat{U}}{d\omega}\right)e^{\omega}. \label{EqsAomega}
\end{eqnarray}
Solutions to these equations can be obtained by imposing the
condition $\omega=2kA$ and following the conformal technique
described in \cite{ariasetal,bh2}. In this work we shall reconsider
the concrete family of solutions
\cite{dewolfe,gremm},\cite{ariasetal}--\cite{bh3}
\begin{eqnarray}
e^{2A(y)}=\left[\cos\left(\sqrt{8\lambda
b}\,(y-c)\right)\right]^{\frac{3}{2b}},\nonumber\\
e^{\omega(y)}=\left[\cos\left(\sqrt{8\lambda
b}\,(y-c)\right)\right]^{-\frac{2}{b}}, \label{sol}
\end{eqnarray}
where $\widehat{U}=\lambda e^{16\xi\omega}$, $b=1+16\xi$, $c$ is an
arbitrary integration constant and $\lambda$ is a coupling constant.
This solution corresponds to the self--interacting potential
$U=\lambda e^{(1+16\xi)\omega}$ in the Weyl frame.

The $\xi$ parameter determines the nature of the scalar field. In
particular, for positive values of $\xi$ ($b>1$), the scalar field
$\omega$ turns out to be a phantom (ghost), as is seen from
(\ref{Eaction}). If $\xi$ is negative ($b<1$), $\omega$ is a
conventional scalar field.


In \cite{bh3} it was shown that for $\lambda,b >0 $ and $b\ne
{15\over 8}$ the 5D curvature scalar $R_5$ is singular at
$\sqrt{8\lambda b}(y-c)=\pm {\pi\over 2}$, so that the domain of the
extra coordinate should be chosen as the interval $-\frac{\pi}{2}\le
\sqrt{8\lambda b}\, (y-c)\le\frac{\pi}{2}$. These curvature
singularities are naked ones, as one can easily show analytically.

\section{
Perturbing the metric}

Following \cite{rs1,dewolfe}, let us study the metric fluctuations
$h_{mn}$ of the metric (\ref{line})  given by the perturbed line
element
\begin{equation}
\label{fluct} ds_5^2=e^{2 A(y)}[\eta_{mn}+h_{mn}(x,y)]dx^m
dx^n+dy^2.
\end{equation}

One must consider the fluctuations of the scalar field when treating
the ones of the metric since they are coupled. However, in
\cite{dewolfe} it was shown that the transverse traceless modes of
the metric fluctuations $h_{mn}^T$ decouple from the scalar sector,
allowing us to study the dynamics of these fluctuations analytically
in closed form. Following \cite{rs2}--\cite{gremm} we perform the
coordinate transformation
\begin {equation}
dw=e^{-A}dy.\label{trafo}
\end {equation}
This leads to a conformally flat metric and to the following wave
equation for the transverse traceless modes of the metric
perturbations,
\begin{equation}
\label{modeequation}
(\partial_w^2+3A'\partial_w+\Box^{\eta})h_{mn}^T=0,
\end{equation}
where $\Box^{\eta}$ is the standard wave operator in four
dimensions. Due to the unbroken Poincar\'e invariance, this equation
supports a massless and normalizable 4D graviton given by
$h_{mn}^T=C_{mn}{\rm e}^{ipx}$, where $C_{mn}$ are constant
parameters and $p^2=m^2=0$. Using the ansatz $h_{mn}^T= {\rm
e}^{ipx}e^{-3A/2}\Psi_{mn}(w)$ the wave equation
(\ref{modeequation}) can be recast in the form of a Schr\"odinger's
equation
\begin{equation}
\label{schroedinger} [\partial_w^2-V(w)+m^2]\Psi=0.
\end{equation}
Here we have dropped the subscripts on $\Psi$, $m$ is the mass of
the KK excitation, the potential reads
\begin{equation}
\label{V(w)}
V(w)=\frac{3}{2}\partial_w^2A+\frac{9}{4}(\partial_wA)^2,
\end{equation}
and is fully defined by the curvature of the 5D space-time.

The question which we wish to study now is, are there values of the
parameter $b$ in the two-parameter family of solutions (\ref{sol})
such that (i) the transformation (\ref{trafo}) is decompactifying,
i.e. maps the compact $y$--interval onto the real $w$--line and (ii)
the resulting Schr\"odinger equation (\ref{schroedinger}) shows a
mass gap between the massless (physical) graviton solution and the
massive KK gravitons.

It is easily seen that (i) requires us to restrict $b \leq 3/4$.
Proceeding to (ii), we face the difficulty that for generic values
of $b$ the differential equation (\ref{trafo}) does not allow us to
construct the function $w(y)$ in explicit form. However, we can
still use (\ref{trafo}) and (\ref{V(w)}) to calculate the potential
as a function of $y$:
\begin{eqnarray}
\label{Vexpl} V(w(y))=9\lambda\Bigl[\cos(\sqrt{8\lambda b}
(y-c))\Bigr]^{\frac{3}{2b}}
\nonumber \\
\times \Bigl[-{\frac{15}{8b}}+\Bigl({\frac{15}{8b}}-1\Bigr)
{\frac{1}{\cos^2(\sqrt{8\lambda b} (y-c))}} \Bigr] \label{Vwy}
\end{eqnarray}

Moreover, the zero-mass solution for the potential (\ref{V(w)}) is
the ground state, and is given simply by \cite{csaki}
\begin{eqnarray}
\Psi_0(w)= C_0 e^{3A(w)/2} \label{Psi0gen}
\end{eqnarray}
as can be verified using (\ref{V(w)}) ($C_0$ is a normalization
constant). From (\ref{trafo}), $w(y)$ is a monotonous function, so
that
\begin{eqnarray}
\lim_{w\to \pm \infty} V(w)=\lim_{\sqrt{8\lambda b}(y-c)\to \pm
{\frac{\pi}{2}}} V(w(y)). \label{limV}
\end{eqnarray}
From (\ref{Vwy}) this limit is zero for $b<3/4$, while for the
borderline case $b=3/4$ it approaches the finite value $V_{\infty}=
27\lambda/2$. This is sufficient to exclude a mass gap for $b<3/4$,
since the Schr\"odinger equation (\ref{schroedinger}) will have a
continuous spectrum starting at zero if the potential goes to zero
at infinity.

Thus, in the following we will focus on the case $b=3/4$. Here
(\ref{trafo}) can be solved explicitly, yielding
\begin{eqnarray}
\cos[\sqrt{6\lambda}(y-c)]={\rm
sech}\bigl[\sqrt{6\lambda}(w-w_0)\bigr]. \label{trafospecial}
\end{eqnarray}
The function $A(w)$ takes the form
\begin{eqnarray}
A(w) = {\rm ln} \Bigl\lbrace {\rm sech} \Bigl[
\sqrt{6\lambda}(w-w_0)\Bigr] \Bigr\rbrace, \label{Aspecial}
\end{eqnarray}
and the potential becomes a modified P\"oschl-Teller one:
\begin{eqnarray}
V(w) =\frac{9\lambda}{2} \Bigl\lbrace 3 - 5\, {\rm sech}^2\Bigl[
\sqrt{6\lambda}(w-w_0)\Bigr] \Bigr\rbrace.
\end{eqnarray}
Since this potential asymptotically reaches a positive value, we
automatically get a mass gap for this case, allowing for a
phenomenological solution of the dangerous light KK excitations
problem (similar results have been obtained for de Sitter
\cite{wang,ps} and non standard \cite{bs} branes). The corresponding
Schr\"odinger equation (\ref{schroedinger}) turns, after a rescaling
$u= \sqrt{6\lambda}(w-w_0)$, into
\begin{eqnarray}
\Bigl[\partial_u^2 + \frac{15}{4}{\rm sech}^2(u)
+\frac{m^2}{6\lambda} - \frac{9}{4}\Bigr] \Psi = 0.
\label{schroedingerspecial}
\end{eqnarray}
In this standardized form it is the special case $n=3/2$ of the
well-known family of Schr\"odinger equations
\begin{eqnarray}
\Bigl[-\partial_u^2  - n(n+1){\rm sech}^2(u)\Bigr] \Psi = E \Psi .
\label{poeschlteller}
\end{eqnarray}
This implies (see, e.g., \cite{poshltellerp}) that in this potential
there are two bound states, the ground state $\Psi_0$ with energy
$E_0\!=\!-n^2\!=\!-9/4$ and one excited state $\Psi_1$ with energy
$E_1\!=\!-(n\!-\!1)^2\!=\!-1/4$. The ground state wave function is
\begin{eqnarray}
\Psi_0(w) = C_0\,{\rm sech}^{3/2}(w). \label{groundstate}
\end{eqnarray}
This is just the zero-mass state (\ref{Psi0gen}), confirming the
absence of tachyonic modes with $m^2 <0$. The wave function of the
excited state is
\begin{eqnarray}
\Psi_1(w) = C_1\, {\rm sinh}(w){\rm sech}^{3/2}(w).
\label{excitedstate}
\end{eqnarray}
It represents a massive graviton of mass $m^2=12\lambda$ localized
on the brane. The continuous spectrum starts at $E=0$, corresponding
to
\begin{eqnarray}
m^2 &\geq & \frac{27}{2} \lambda. \label{contspec}
\end{eqnarray}
These states asymptotically turn into plane waves, and represent
delocalized KK massive gravitons. Their explicit expressions can be
given in terms of the associated Legendre functions of the first
kind $P^{\mu}_{\nu}$, of degree $\nu = 3/2$ and purely imaginary
order $\mu=i\rho$:
\begin{eqnarray}
\Psi^{\mu}(w) &=& \sum_{\alpha= \pm} C_{\alpha}\, P^{\alpha
i\rho}_{\frac{3}{2}} \Bigl({\rm tanh}\Bigl[
\sqrt{6\lambda}(w-w_0)\Bigr]\Bigr) \label{massivemodes}
\end{eqnarray}
where $C_{\pm}$ are integration constants and
\begin{eqnarray}
\rho=\vert{\mu}\vert=\sqrt{\frac{m^2}{6\lambda}-\frac{27}{12}}.
\label{defrho}
\end{eqnarray}
It should be mentioned that the wave functions of these KK gravitons
are, in the original $y$ coordinates, singular at the endpoints $y =
c \pm {\pi\over 2 \sqrt{6\lambda}}$.

To study the brane content of the $b=3/4$ solution we calculate the
energy density. In the $w$ coordinates this is

\begin{eqnarray}
 T_{00}(w)&=&-{3\over 8\pi G_5}\left[\partial_w^2A+(\partial_wA)^2\right] \nonumber\\
 &=& {9\lambda\over 4\pi G_5}\biggl({2\over\cosh^2(\sqrt{6\lambda}\, w)}-1\biggr) \nonumber\\
 \label{T00}
\end{eqnarray}
In the limit of large $\lambda$ this becomes

\begin{equation}
T_{00} \simeq {9\over\pi G_5} \Bigl(\sqrt{{\lambda \over 6}}
\delta(w)- {\lambda\over 4} \Bigr) \label{T00asymp}
\end{equation}
Thus, up to a cosmological constant term, the energy density
localizes at a single brane located at the center, $w=0$. For the
applications presented below the constant term is not relevant,
since it can be removed by a renormalization of the vacuum energy.
 Finally, it should be mentioned that in the $w$ coordinates the curvature singularities
are at $w=\pm \infty$, but still at a finite geodesic distance from
the brane.

\section{
Physical applications}

To apply our brane configuration in the context of
\cite{rs1,rs2,lr}, first we need to establish the connection between
the Planck scales in four and five dimensions. In order to get the
four-dimensional effective Planck scale, we replace the Minkowski
metric by a four-dimensional metric $\overline{g}_{mn}$ in eq.
(\ref{line}), leading to an effective four-dimensional action. We
look at the curvature term from which one can derive the scale of
the gravitational interactions
\begin{equation}
\label{actioneffective} S_{eff}\supset\int d^4x \int_{0}^{\infty}
dw\, 2M_{\ast}^3
\sqrt{|\overline{g}|}e^{\frac{3}{2}\omega}e^{3A}\overline{R},
\end{equation}
where we performed the coordinate transformation (\ref{trafo}),
$M_{\ast}$ is the Planck scale in five dimensions, and
$\overline{R}$ is the four-dimensional Ricci scalar. Because the low
energy fluctuations have no $w$ dependence, we can perform the
direct integration along $w$, obtaining a purely four-dimensional
action, and calculate the four-dimensional effective Planck scale
\begin{equation}
 M_{pl}^2=2M_{\ast}^3\int_{0}^{\infty}dw\, e^{\frac{3}{2}\omega}e^{3A}.
\end{equation}
For our particular solution, eq. (\ref{sol}) with $b=3/4$, we get
for large $r$
\begin{equation}
\int_{0}^r dw\, e^{\frac{3}{2}\omega}e^{3A} \approx
\frac{1}{\sqrt{6\lambda}}\left(\frac{5\pi}{32}-\frac{128}{7}e^{-7\sqrt{6\lambda}r}\right),
\end{equation}
so that
\begin{equation}
M_{pl}^2=\frac{5\pi}{16\sqrt{6\lambda}}M_{\ast}^3. \label{Mpl}
\end{equation}
Thus $M_{pl}$ is finite, and we also see that it depends only weakly
on $r$ for $r\rightarrow\infty$.

\subsection{
KK corrections to Newton's law}

The massive gravitons give rise to corrections to Newton's law in
ordinary 4D flat spacetime. In the present model and in the thin
brane limit, they can be expressed in the following way
\cite{rs2,csaki}:
\begin{eqnarray}
U(r)\sim \frac{M_1M_2}{r} \biggl[
G_4 + M_{\ast}^{-3} e^{-m_1r} \vert\hat\Psi_1(w_0)\vert^2 \nonumber \\
+ M_{\ast}^{-3}\int_{m_0}^{\infty}dm\, e^{-mr}
\vert{\hat\Psi^{\mu(m)}(w_0)}\vert^2 \biggr]. \label{U}
\end{eqnarray}
Here the thin brane represents the physical universe, located at
$w=w_0$ in the extra dimension. $G_4$ denotes the four dimensional
gravitational coupling, $\hat\Psi_1$ is the wave function of our
single normalizable excited state, and $\hat\Psi^{\mu(m)}$ denotes
the continuous eigenfunctions which have to be integrated over their
masses, and (except for $\mu=0$) to be summed over the two
independent eigenfunctions. From (\ref{massivemodes}) those can be
chosen as (setting $w_0=0$ for convenience)
\begin{eqnarray}
\hat\Psi^{\mu}_{\pm}(w)= C_{\pm}(\rho)P^{\pm i\rho }_{\frac{3}{2}}
\Bigl(\tanh\bigl(\sqrt{6\lambda}\, w\bigr)\Bigr). \label{Psipm}
\end{eqnarray}
The `hat' on $\hat\Psi_1$ means that it is normalized, while for the
$\Psi^{\mu}$, which asymptotically for large $\vert w \vert$ are of
plane wave form, it means normalization in the plane wave sense. The
thin brane limit has been used to set both test bodies to the center
of the brane in the transverse direction, $w=0$. Since
$\hat\Psi_1(w)$ is an odd function it does not contribute in this
limit. To calculate the corrections due to the continuous modes,
first we have to compute the normalization constants
$C_{\pm}(\rho)$. Using the large $x$ approximation of the
$\tanh(x)$,
\begin{equation}
\tanh(x) \simeq 1 - 2e^{-2x} \label{asymptanh}
\end{equation}
and Eq. (8) of section 3.9.2 of \cite{erdelyi}, one can easily show
that for $w\to\infty$
\begin{eqnarray}
P^{\pm i\rho}_{\frac{3}{2}}\Bigl(\tanh\bigl(\sqrt{6\lambda}
w\bigr)\Bigr) \sim \frac{1}{\Gamma(1\mp i\rho)}}e^{\pm
i\sqrt{6\lambda} \rho w. \label{Pasymp}
\end{eqnarray}
Since $\vert{\Gamma(1-i\rho)}\vert=\vert{\Gamma(1+i\rho)}\vert$ this
yields
\begin{eqnarray}
C_+(\rho)=C_-(\rho) =\frac{\vert\Gamma(1+i\rho)\vert}{\sqrt{2\pi}}
\label{Cpmfin}
\end{eqnarray}
Further, we need the value of $\vert{\hat\Psi^{\mu}(0)}\vert$. From
\cite{erdelyi}
\begin{eqnarray}
P^{\mu}_{\nu}(0)=\frac{2^{\mu}}{\sqrt{\pi}}
\cos\Bigl[\frac{1}{2}\pi(\nu+\mu)\Bigr]\frac{\Gamma\Bigl(\frac{1}{2}+\frac{1}{2}\nu+
\frac{1}{2}\mu\Bigr)}{\Gamma\Bigl(1+\frac{1}{2}\nu-\frac{1}{2}\mu\Bigr)}
\label{P0}
\end{eqnarray}
For $\mu = \pm i\rho$ and taking the absolute value this can be
simplified to
\begin{eqnarray}
\bigr\vert{P^{\pm
i\rho}_{\frac{3}{2}}(0)}\bigr\vert=\frac{\sqrt{\pi}}
{\Bigr\vert\Gamma\Bigl(-\frac{1}{4}+\frac{i\rho}{2}\Bigr)\Bigr\vert
\Bigr\vert\Gamma\Bigl(\frac{7}{4} +
\frac{i\rho}{2}\Bigr)\Bigr\vert}. \label{absP0}
\end{eqnarray}
Putting things together, we can express the content of the square
brackets in (\ref{U}) as $G_4 + \Delta G_4$, where
\begin{eqnarray}
\Delta G_4\!=\!M_{\ast}^{-3}\!\int_{m_0}^{\infty}\!dm\,e^{-mr}
\Bigr\vert\frac{\Gamma\Bigl(1+i\rho\Bigr)}{\Gamma\Bigl(\!-\frac{1}{4}\!+\!\frac{i\rho}{2}\Bigr)
\Gamma\Bigl(\frac{7}{4}\!+\!\frac{i\rho}{2}\Bigr)}\Bigr\vert^2\!.
\label{squarebrack}
\end{eqnarray}
For the calculation of this integral, it will be convenient to
change variables from $m$ to $\rho$. This yields
\begin{eqnarray}
\Delta G_4=M_{\ast}^{-3}\sqrt{6\lambda}
\int_{0}^{\infty}\frac{d\rho}{\sqrt{1+\frac{27}{12\rho^2}}}
e^{-r\sqrt{6\lambda}\sqrt{\rho^2+\frac{27}{12}}}\nonumber\\
\times
\Bigr\vert\frac{\Gamma\Bigl(1+i\rho\Bigr)}{\Gamma\Bigl(-\frac{1}{4}+\frac{i\rho}{2}\Bigr)
\Gamma\Bigl(\frac{7}{4} + \frac{i\rho}{2}\Bigr)}\Bigr\vert^2.
\label{squarebrackfin}
\end{eqnarray}
Although it seems not possible to do this integral in closed form,
it can be easily calculated numerically as a function of
$\sqrt{6\lambda}\, r$. Moreover, in the thin brane limit
$\lambda\to\infty$ the integral will be dominated by the small -
$\rho$ region, so that it can be well-approximated by expanding the
prefactor of the exponential at $\rho =0$. In this way one obtains
for the thin brane limit
\begin{eqnarray}
\Delta G_4 \sim&M_{\ast}^{-3}
\frac{1}{\Bigr\vert\Gamma(-\frac{1}{4})
\Gamma(\frac{7}{4})\Bigr\vert^2}
\frac{e^{-\sqrt{\frac{27}{2}\lambda} r}}{r} (1+
O(\frac{1}{r\sqrt{\lambda}})). \label{approxDeltaGN}
\end{eqnarray}
We remark that, away from the thin brane limit, the correction would
also involve the massive bound state. The form of Newton's law would
then become rather involved, and depend on the precise location of
the two test bodies along the fifth dimension. Further corrections
to Newton's law might be induced by the scalar field modes, whose
spectrum we have not attempted to analyze here. Corrections to
Newton's law were also obtained for brane models with a
non-minimally coupled bulk scalar field in \cite{farakosetal}.

%
\subsection{
Solution of the mass hierarchy problem}

Our special configuration also fits well into the framework of
\cite{lr}, where it was shown how to arrive at a solution of the
mass hierarchy problem using a non-compact extra dimension. First,
we observe that the warp factor (\ref{Aspecial}) in the stiff limit
reproduces the Randall--Sundrum one:
\begin{equation}
e^{2A}\simeq 4 e^{-2\sqrt{6\lambda}\vert w \vert }. \label{rsfactor}
\end{equation}
In the set-up of \cite{lr}, the limiting thin brane at $w=0$ does
not represent the physical brane, but a hidden ``Planck brane''. The
physical brane will be located at a distance $w$, where the
effective Planck scale is reduced from $M_{Pl}=10^{19}GeV$ by the
factor (\ref{rsfactor}). By an appropriate choice of $w$ this
exponential factor can be made to account for the unwanted
discrepancy between the Planck and electroweak scales. Using our
special brane configuration, which has a mass gap proportional to
$\sqrt{\lambda}$, we can implement this solution of the hierarchy
problem without having to face possible problems with low--mass KK
graviton excitations. However, this issue will be studied in more
detail in \cite{bhnq} in a Riemannian manifold.

\vspace{20pt}

\section{
Concluding remarks}

Continuing previous work by two of the authors \cite{bh3} on Weyl
thick branes, we have performed an analysis of the transverse
traceless modes of the linear fluctuations of a family of classical
backgrounds. For a particular case, we have arrived at an exact
solution of the effective Schr\"odinger equation obeyed by these
modes. This is exceptional, since usually for smooth brane
configurations with Poincar\'e symmetry one is not able to integrate
the Schr\"odinger equation for the massive modes explicitly (a toy
model with calculable KK modes was given in \cite{csaki}). More
importantly, the resulting graviton spectrum is unusual in that it
shows both a mass gap and a continuum of massive KK modes, as well
as a single massive localized graviton state. The mass gap is
proportional to the scalar self-interaction coupling constant
$\sqrt{\lambda}$. Its existence provides an easy way to control the
excitation of the KK gravitons and concomitant energy loss into the
fifth dimension. This suppression of the delocalized gravitons
should also render harmless the presence of naked singularities at
$w= \pm\infty$. We have given two applications: using this
configuration along the lines of RS \cite{rs2} we have, in the thin
brane limit, obtained the correction to Newton's law in closed form;
only the massive modes contribute in this limit. In the framework of
\cite{lr}, our configuration leads to a solution of the hierarchy
problem without low-energy KK graviton excitations.

It is worth noticing that one can start with a setup free of phantom
fields in the Einstein frame from the very beginning as in
\cite{dewolfe,gremm,bgl}, consider solutions for the warp factor of
the form $e^{2A}=\cos^b(ay)$ (with a suitable self--interacting
potential) and properly set the value for the parameter $b=2$ in
order to ensure the appearance of the mass gap \cite{bhnq}. To the
best of our knowledge, only the non--compact version of this
solution $e^{2A}=\cosh^b(ay)$ has been studied in the literature.

\section{Acknowledgements}

AHA thanks the organizers of the {\it 2nd International Meeting on
Gravitation and Cosmology} held at Universidad de Las Villas for
their hospitality, and the participants for fruitful discussions,
specially with C. Germani, R. Maartens and U. Nucamendi. CS thanks
O. Corradini for discussions. NBC acknowledges a PhD grant from
CONACYT. Finantial assistance under grants No. CONACYT-J60060,
COECYT-CIC-4.16 and CONACYT-SEP-2003-C02-45364 is acknowledged by
AHA and MARS, respectively.


\begin{thebibliography}{99}

\bibitem{rubakov} K. Akama, Lec. Notes Phys {\bf 176}, 267 (1980);
V.A. Rubakov and M.E. Shaposhnikov, Phys. Lett. B{\bf 125}, 139
(1983); M. Visser, Phys. Lett. B{\bf 159}, 22 (1985); E.J. Squires,
Phys. Lett. B{\bf 167}, 286 (1986); A. Barnaveli and O. Kancheli,
Sov. J. Nucl. Phys. {\bf 51}, 901 (1990); Sov. J. Nucl. Phys. {\bf
52}, 920 (1990); I. Antoniadis, Phys. Lett. B{\bf 246}, 317 (1990).

\bibitem{ahddd} N. Arkani-Hamed, S. Dimopoulos and G. Dvali, Phys. Lett. B{\bf 429}, 263 (1998).

\bibitem{aahddd} I. Antoniadis, N. Arkani-Hamed, S. Dimopoulos and G. Dvali, Phys. Lett. B{\bf 436}, 257 (1998).

\bibitem{gog} M. Gogberashvili, Mod. Phys. Lett. A{\bf 14}, 2025 (1999);
Int. J. Mod. Phys. D{\bf 11}, 1635 (2002); Europhys. Lett. {\bf 49},
396 (2000).

\bibitem{rs1} L. Randall and R. Sundrum, Phys. Rev Lett. {\bf 83}, 3370 (1999).

\bibitem{rs2} L. Randall and R. Sundrum, Phys. Rev. Lett. {\bf 83}, 4690 (1999).

\bibitem{lr} J. Lykken and L. Randall, JHEP {\bf 0006}, 014 (2000).

\bibitem{dewolfe} O. De Wolfe, D. Z. Freedman, S. S. Gubser and A. Karch, Phys. Rev. D{\bf 62}, 046008 (2000).

\bibitem{gremm} M. Gremm, Phys. Lett. B{\bf 478}, 434 (2000); Phys. Rev. D{\bf 62}, 044017 (2000).

\bibitem{scalarthickbranes} A. Melfo, N. Pantoja and A. Skirzewski,
Phys. Rev. D{\bf 67}, 105003 (2003); C.--J. Zhu, JHEP {\bf 0006}, 34
(2000); R. Guerrero, A. Melfo and N. Pantoja, Phys. Rev. D{\bf 65},
125010 (2002); R. Koley and S. Kar, Class Quantum Grav. {\bf 22},
753 (2005); D. Bazeia and L. Losano, Phys. Rev. D{\bf 73}, 025016
(2006); V. Dzhunushaliev, Grav. Cosmol. {\bf 13}, 302 (2007); S.
Ghassemi, S. Khakshournia and R. Mansouri, JHEP {\bf 0608}, 019
(2006); Int. J. Mod. Phys. D{\bf 16}, 629 (2007); A.A. Saharian,
A.L. Mkhitaryan, JHEP {\bf 0708}, 063 (2007); V. Dzhunushaliev, V.
Folomeev, K. Myrzakulov and R. Myrzakulov, ``Thick brane in 7D and
8D spacetimes", arXiv:0705.4014 [gr-qc]; N. Barbosa-Cendejas and A.
Herrrera-Aguilar, J. Phys. Conf. Ser. {\bf 68}, 012021 (2007).

\bibitem{branetachyons} D. Bazeia, F.A. Brito and J.R. Nascimento, Phys. Rev. D{\bf 68}, 085007 (2003);
R. Koley and S. Kar, Phys. Lett. B {\bf 623}, 244 (2005).

\bibitem{BlackHoles} R. Emparan, R. Gregory and C. Santos, Phys. Rev. D{\bf 63}, 104022 (2001);
N. Dadhich, R. Maartens, P. Papadopoulos, V. Rezania, Phys. Lett.
B{\bf 487}, 1 (2000).

\bibitem{wang} A. Wang, Phys. Rev. D{\bf 66}, 024024 (2002).

\bibitem {griegos} C. Bogdanos, A. Dimitriadis, and K. Tamvakis, Class. Quant. Grav. {\bf 24}, 3701 (2007);
K. Farakos, P. Pasipoularides, J. Phys. Conf. Ser. {\bf 68} 012041
(2007); C. Bogdanos, J. Phys. Conf. Ser. {\bf 68}, 012045 (2007).

\bibitem {farakosetal} K. Farakos, G. Koutsoumbas and P. Pasipoularides, Phys. Rev. D {\bf76}, 064025 (2007).

\bibitem{csaki} C. Csaki, J. Erlich, T. Hollowood and Y. Shirman, Nucl. Phys. B{\bf 581}, 309 (2000).

\bibitem{Localgravity1} J. Garriga and T. Tanaka, Phys. Rev. Lett. {\bf 84}, 2778 (2000);
A. Kehagias and K. Tamvakis, Phys. Lett. B{\bf 504}, 38 (2001); R.
Guerrero, R. Ortiz, R. O. Rodriguez and R. S. Torrealba Gen. Rel.
Grav. {\bf 38}, 845 (2006); R. Maartens, Living Rev. Rel. {\bf 7},
07 (2004); L.A. Gergely and R. Maartens, Phys. Rev. D {\bf71},
024032 (2005).

\bibitem{Localgravity2} D. Bazeia, F.A. Brito and A.R. Gomes, JHEP {\bf 0411}, 070 (2004); O.
Castillo--Felisola, A. Melfo, N. Pantoja and A. Ramirez, Phys. Rev.
D{\bf 70}, 104029 (2004).

\bibitem{ariasetal} O. Arias, R. Cardenas and Israel Quiros, Nucl. Phys. B{\bf 643}, 187 (2002).

\bibitem{bh2} N. Barbosa--Cendejas and A. Herrera--Aguilar, JHEP {\bf 0510}, 101 (2005).

\bibitem{bh3} N. Barbosa--Cendejas and A. Herrera--Aguilar, Phys. Rev. D {\bf73}, 084022 (2006);
Erratum-ibid. {\bf 77}, 049901 (2008).

\bibitem{bgl} D. Bazeia, A.R. Gomes and L. Losano, "Gravity localization on thick branes: a numerical approach", arXiv:0708.3530 [hep-th].

\bibitem{matterloc} Y.-X. Liu, X.-H. Zhang, L.-D. Zhang, and Y.-S. Duan,
JHEP {\bf0802}, 067 (2008); X.-H. Zhang, Y.-X. Liu and Y.-S. Duan,
``Localization of fermionic fields on braneworlds with bulk tachyon
matter", arXiv:0709.1888 [hep-th] .

\bibitem{stabilitythickbranes1} S. Kobayashi, K. Koyama and J. Soda, Phys. Rev.
D{\bf 65}, 064014 (2002); M. Minamitsuji, W. Naylor and M. Sasaki,
Phys. Lett. B{\bf 633}, 607 (2006).

\bibitem{stabilitythickbranes2} S. Kobayashi, K. Koyama and J.
Soda, Phys. Lett. B {\bf501}, 157 (2001); E. Abdalla, B.
Cuadros--Melgar, A.B. Pavan, and C. Molina, J. Phys. Conf. Ser. {\bf
68}, 012043 (2007).

\bibitem{criticalphenomena} A. Campos, Phys. Rev. Lett. {\bf 88}, 141602 (2002).

\bibitem{ps} M.K. Parikh and S.N. Solodukhin, Phys. Lett. B{\bf 503}, 384 (2001).

\bibitem{bs} A. Brandhuber and K. Sfetsos, JHEP {\bf 9910}, 013 (1999).

\bibitem{poshltellerp} J.I. Diaz, J. Negro, L.M. Nieto, O. Rosas--Ortiz,
J. Phys. A: Math.Gen. {\bf32}, 8447 (1999).

\bibitem{erdelyi}
A. Erd\'elyi (Ed.), {\sl Higher transcendental functions, Vol. 1},
McGraw-Hill 1953, repr. R.E. Krieger Publ. 1981.

\bibitem{bhnq} N. Barbosa--Cendejas, A. Herrera--Aguilar, U. Nucamendi and I. Quiros,
arXiv:0712[hep-th].


\end{thebibliography}
\end{document}